\documentclass{aastex631}
\usepackage{graphicx}
\usepackage{amsmath}
\usepackage{hyperref}

\usepackage{xcolor}
\usepackage{soul}

\begin{document}

\title{Multi-wavelength JWST observations of (3200) Phaethon show a dehydrated object with an aqueously altered origin}

\author[0000-0003-3091-5757]{Cristina A. Thomas}
\affiliation{Northern Arizona University, Department of Astronomy and Planetary Science,
P.O. Box 6010,
Flagstaff, AZ, 86011 USA}

\author{Andrew S. Rivkin}
\affiliation{Johns Hopkins University Applied Physics Laboratory, 11100 Johns Hopkins Road,
Laurel, MD, 20723 USA}

\author[0000-0001-9665-8429]{Ian~Wong}
\affiliation{Space Telescope Science Institute, 3700 San Martin Drive, Baltimore, MD 21218, USA}
\affiliation{NASA Goddard Space Flight Center, 8800 Greenbelt Road, Greenbelt, MD 20771, USA}

\author[0000-0003-2781-6897]{Matthew M. Knight}
\affiliation{Physics Department, United States Naval Academy, 572C Holloway Rd, Annapolis, MD 21402, USA}

\author[0000-0002-8144-7570]{Sean E. Marshall}
\affiliation{Florida Space Institute, University of Central Florida, 12354 Research Pkwy, Orlando, FL, 32826, USA}

\author{Christopher W. Haberle}
\affiliation{Northern Arizona University, Department of Astronomy and Planetary Science,
P.O. Box 6010,
Flagstaff, AZ, 86011 USA}

\author{Aidan Madden-Watson}
\affiliation{Northern Arizona University, Department of Astronomy and Planetary Science,
P.O. Box 6010,
Flagstaff, AZ, 86011 USA}

\author{Joshua P. Emery}
\affiliation{Northern Arizona University, Department of Astronomy and Planetary Science,
P.O. Box 6010,
Flagstaff, AZ, 86011 USA}

\author[0000-0002-7600-4652]{Annika Gustafsson}
\affiliation{Northern Arizona University, Department of Astronomy and Planetary Science,
P.O. Box 6010,
Flagstaff, AZ, 86011 USA}

\author [0000-0001-7694-4129]{Stefanie N. Milam}
\affiliation{NASA Goddard Space Flight Center, 8800 Greenbelt Road, Greenbelt, MD 20771, USA}

\author{Heidi B. Hammel}
\affiliation{Association of Universities for Research in Astronomy, Washington, DC, USA}

\author{Ellen S. Howell}
\affiliation{University of Arizona, Lunar and Planetary Laboratory}

\author[0000-0002-8227-9564]{Ronald J. Vervack Jr.}
\affiliation{Johns Hopkins University Applied Physics Laboratory, 11100 Johns Hopkins Road,
Laurel, MD, 20723 USA}



\begin{abstract}

We present JWST observations of the near-Earth asteroid (3200) Phaethon using the Near-Infrared Camera (NIRCam), Near-Infrared Spectrograph (NIRSpec), and Mid-Infrared Instrument (MIRI) to further investigate the composition of Phaethon's surface. Our NIRSpec data confirms that Phaethon's surface is dehydrated, showing no evidence of hydrated minerals in the 3-$\mu$m region. We estimate an upper limit on the hydrogen content in phyllosilicates of 0.06 wt\%. Comparisons with laboratory spectra of carbonaceous chondrites suggest that Phaethon's surface composition is best matched by thermally metamorphosed samples of the CM chondrite Murchison (heated to 1000$^{\circ}$C), rather than CY meteorites as previous work suggested. We find no evidence of ongoing surface evolution due to recent perihelion passages. A comparison of the mid-infrared spectra of Phaethon and Bennu shows distinct spectral differences that are consistent with their different thermal histories. Our findings further refine our understanding of Phaethon's current surface composition and evolution and provide additional insights for the upcoming DESTINY\textsuperscript{+} mission.

\end{abstract}



\section{Introduction} \label{sec:intro}

Since its discovery in 1983 \citep{green1983}, near-Earth asteroid (3200) Phaethon has been an intriguing target for study. It is the second largest potentially hazardous asteroid, and its combination of a very low perihelion distance ($q=0.140$~au)\footnote{All orbital information in this paper is from \href{https://ssd.jpl.nasa.gov/horizons/}{JPL Horizons}} and a large size (diameter of 4.6, 5.1, or 6.2~km; \citet{masiero2019thermophysical}, \citet{hanus2016}, \citet{taylor2019}, respectively) 
makes it stand out among the near-Earth asteroid population. It has a short orbital period (1.43~years) and a small minimum orbit intersection distance from Earth (MOID = 0.019~au), resulting in many favorable opportunities for observations. Especially close approaches to Earth in 2007 (minimum geocentric distance, $\Delta_\mathrm{min}$, of 0.121~au) and 2017 ($\Delta_\mathrm{min} = 0.069$~au) yielded a variety of studies, including phase-resolved spectroscopy \citep[e.g.,][]{kareta2018,ohtsuka2020}, extensive lightcurves \citep[e.g.,][]{hanus2016,maclennan2022surfacehetero}, radar detection \citep{taylor2019}, and radar-derived shape modeling \citep{2024DPS_Marshall, Marshall2025Phaethon}.

Based upon orbital similarities, Phaethon is thought to be the parent of the Geminid meteor shower \citep[e.g.,][]{whipple1983,williams1993}, but various investigations have failed to detect activity sufficient to support the current Geminid flux. Observations by solar observatories detected activity near perihelion on repeated orbits \citep[e.g.,][]{jewitt2010,li2013}, but \citet{zhang2023} demonstrated that the near-Sun brightness is likely dominated by resonant fluorescence of sodium atoms, requiring substantially less mass loss than earlier estimates. Observations at larger heliocentric distances, notably around the 2017 close approach, have set upper limits on dust production, cometary volatile outgassing, and the presence of large fragments in its vicinity \citep[e.g.,][]{jewitt2019,tabeshian2019,ye2021}.
Thus, while it is assumed that Phaethon's surface must be losing material due to the extreme perihelion temperatures that exceed 1000~K \citep[e.g.,][]{maclennan2021dynamical}, the depth of material lost remains uncertain, as does the amount of regolith stirring that could be occurring. These assumed surface changes could be observable over time.

Phaethon's B-type classification \citep[e.g.,][]{tedesco2002IRAS,demeo2009extension,deleon2010origin,kareta2018} has long associated it with other primitive asteroids and potential meteorite analogs. Using visible and near-infrared (VNIR) wavelength spectra \cite{licandro2007nature} found that the object was most similar to aqueously altered CI/CM meteorites. The VNIR spectral shape and resulting analyses over the past decades have consistently found the object to be blue-sloped with remarkably little variation \citep[e.g.,][]{kareta2018}. The spectral similarity between Phaethon, Pallas, and members of the Pallas family motivated the search for dynamical links between Pallas and Phaethon's current near-Earth orbit. \cite{deleon2010origin} and \cite{todorovic2018dynamical} used numerical simulations to demonstrate that Phaethon could be from the Pallas family. A more recent analysis by \cite{maclennan2021dynamical} concluded that Pallas is not the most likely parent and suggests an inner Main Belt source, such as the Svea or Polana families.

Rotationally resolved 3-$\mu$m region spectra of Phaethon \citep{takir2020} showed no indication of hydration on the surface, suggesting that surface volatile sublimation was not the source of the observed activity. The nature of Phaethon's surface was further examined by \cite{maclennan2024}. Their analysis of the Spitzer IRS spectrum from January 2005 led to the conclusion that the best meteorite analog for Phaethon is Y-82162 from the carbonaceous chondrite Yamato-group (CY). The ``Belgica Grouplet" of carbonaceous chondrites, which includes Y-82162 (as well as Y-86720 and B-7904), experienced aqueous alteration of their parent bodies and have compositions between CI and CM meteorites \citep{king2019yamato}. These objects were subsequently subjected to late-stage thermal metamorphism (T$>$500$^{\circ}$C) where high temperatures caused mineral decomposition and dehydration \citep[e.g.,][]{king2019yamato,suttle2023mineralogy}. \cite{maclennan2024} used a model of heat diffusion to study the thermal decomposition of Phaethon's surface and sub-surface. They concluded that the surface was dehydrated, but the interior consists of relatively unheated, hydrated CY ``precursor" material. The heating of these buried phyllosilicates every perihelion could produce gas pressure sufficient to lift regolith from the surface. This sequence of processing is consistent with a low-albedo, B-type asteroid that has a history of heating up to $\sim$1050~K \citep{maclennan2021dynamical} during low perihelion ($q=0.14$~au) passages.


We obtained observations Phaethon using JWST \citep[e.g.,][]{Rigby_2023,Gardner_2023}. To fully characterize our target, we used the Near-Infrared Spectrograph \citep[NIRSpec;][]{Boker_2023}, the Near-Infrared Camera \citep[NIRCam;][]{Rieke_2023}, and the Mid-Infrared Instrument \citep[MIRI;][]{Wright_2023}, which together provide continuous wavelength coverage from 0.6 to 28 $\mu$m. The JWST data presented in this article were obtained from the Mikulski Archive for Space Telescopes (MAST) at the Space Telescope Science Institute. The specific observations analyzed can be accessed via \dataset[doi: 10.17909/rrkh-p335]{https://doi.org/10.17909/rrkh-p335}. In this paper, we discuss the observations with a focus on the spectroscopic datasets, compare our results to previous observations, and consider the implications of our findings for the field's current understanding of Phaethon.

\section{Observations} \label{sec:obs}

Phaethon was observed during two visits by JWST as part of Cycle 1 Guaranteed Time Observations program 1245 (PI: Cristina Thomas). The details of the observations, including total effective exposure times and target positions, are provided in Table \ref{table:obs}. JWST observations require the elongation angle of the target to be between 85 and 135$^{\circ}$. During Cycle 1, a maximum limit on the non-sidereal tracking rate of 30 mas/sec was implemented. This maximum rate has since been increased to 75 mas/sec. These requirements limited observations of Phaethon to near aphelion, when it was not expected to be active. 

Phaethon was observed with NIRSpec and NIRCam on 2022 October 11, when it was at $r_h$ (solar distance) of 2.10 au. The NIRSpec observations used the fixed slit mode with the PRISM grating and the NRSRAPID readout pattern to obtain two exposures with effective exposure times of 93.5 s and a total effective exposure time of 187.0 s. Between exposures, the object was dithered along the S200A1 slit.  The PRISM data have continuous spectral coverage across the 0.6--5.3 $\mu$m region at a spectral resolving power of $\lambda/\Delta\lambda \sim 100$. The NIRCam observations immediately followed the NIRSpec sequence. NIRCam obtained simultaneous images of Phaethon in the F090W and F430M filters, which have pivot wavelengths of 0.901 and 4.280 $\mu$m and pixel scales of 0.031 and 0.063 {\arcsec}/px, respectively. We used the 4-point subarray dither pattern with the RAPID readout pattern. Each individual exposure consisted of four integrations and had an effective exposure time of 117.2 s. The total exposure time across the four dithers was 468.8 s.

\begin{figure}[ht!]
\epsscale{0.6}
\plotone{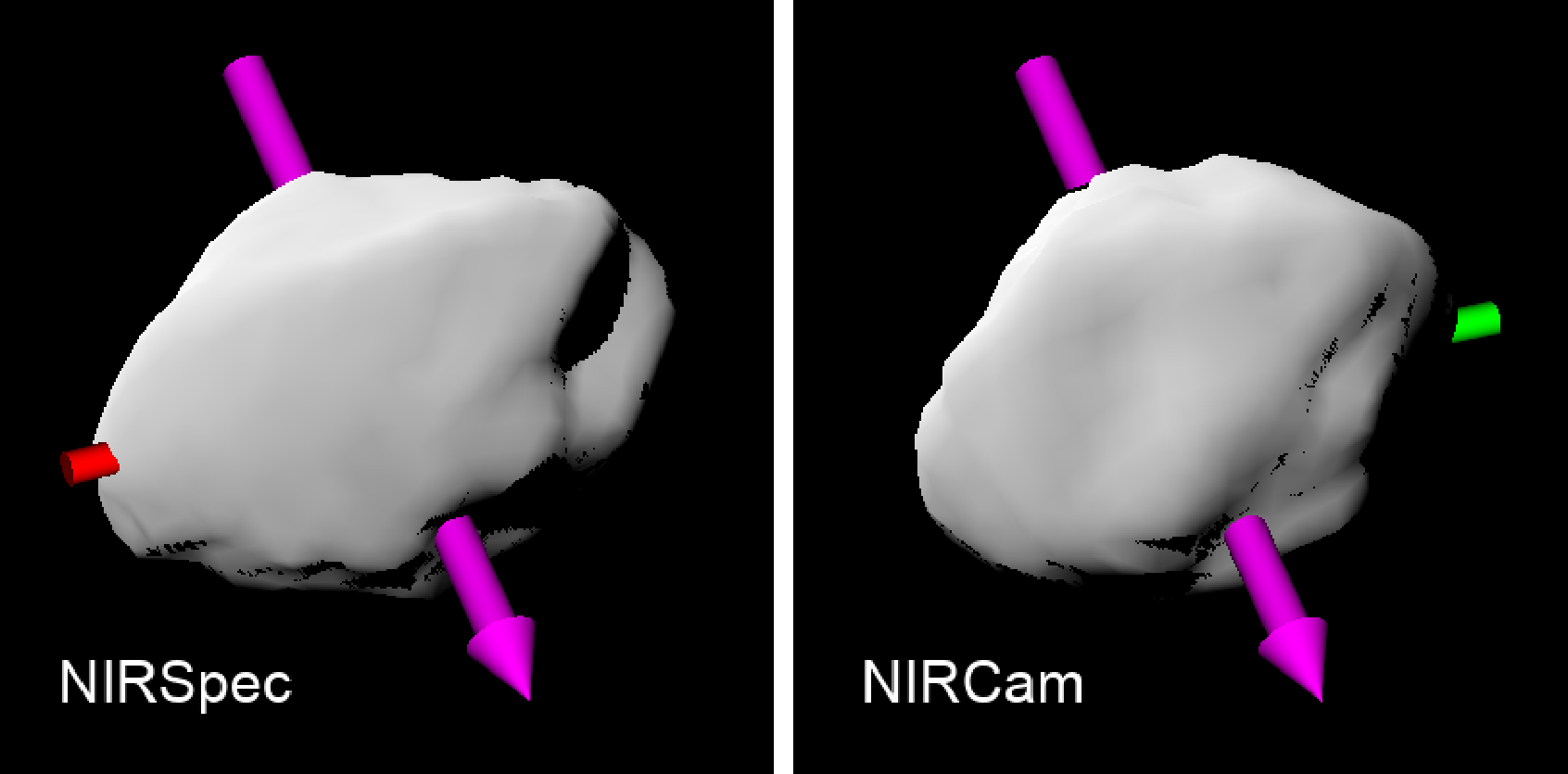}
\caption{Simulated views of Phaethon as seen from JWST during the NIRSpec and NIRCam observations \citep{Marshall2025Phaethon}. In these plane-of-sky views, north is up and east is to the left. The magenta arrow is the spin axis (as defined by the right-hand rule; shortest principal axis) with the arrow pointing to the north pole, and the red and green shafts are the long and intermediate principal axes, respectively.
\label{fig:geom_oct}}
\end{figure}

MIRI Medium Resolution Spectroscopy (MRS) mode observations were obtained on 2022 December 10. MIRI MRS \citep{Wells_2015,argyriou2023} uses four integral field units (IFUs), corresponding to spectral channels 1--4, to provide continuous wavelength coverage between 4.9 and 27.9 $\mu$m across two detectors. Observations are obtained in all four spectral channels simultaneously in one of three grating settings (short, medium, long), with each setting sampling a different sub-band within a given channel's wavelength range. The spectral resolution of the instrument ranges from $\sim$3,500 at 5 $\mu$m to $\sim$1,500 at 28 $\mu$m. Our data were taken with all three grating settings in the long--medium--short sequence over a period of approximately 1.8 h, resulting in spectra that span the full MIRI MRS wavelength range. We used a 4-point dither sequence and the FASTR1 readout pattern with individual effective exposure times of 352.4 s and a total effective exposure time of 1409.7 s per grating setting.

\begin{figure}[ht!]
\plotone{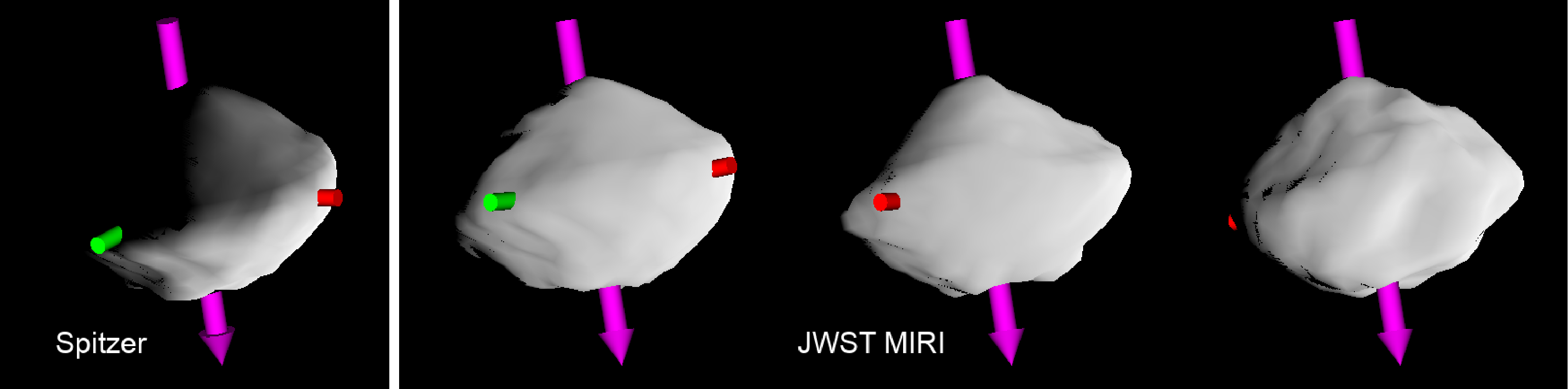}
\caption{Simulated views of Phaethon as seen from Spitzer and JWST during the mid-infrared spectral observations . The plane-of-sky orientations and the shape model are the same as in Figure \ref{fig:geom_oct}. The three MIRI views are at the start of the MIRI observation, mid-observation, and at the end of the observation sequence. 
\label{fig:geom_dec}}
\end{figure}

Phaethon has a rotation period of 3.604~h \citep{hanus2016}, and its rotation period is decreasing by about 4~ms per year \citep{Marshall2022DPS}. For the shape model of Phaethon, \citet{2024DPS_Marshall} and \citet{Marshall2025Phaethon} used the SHAPE software \citep{Magri2007} with radar \citep{taylor2019}, occultation \citep[and others]{Yoshida2023occultation}, and lightcurve \citep[and others]{hanus2016} observations from 1989 through 2022. With that extensive set of observations, it is possible to determine a unique rotation state, with no ambiguity in the number of rotations between apparitions, so we know the sub-observer longitude and latitude during each of the JWST observations to within a few degrees. The stated uncertainty is due to the uncertainty in Phaethon's pole direction. Phaethon is approximately spheroidal with an equatorial ridge, and its volume-equivalent diameter is 5.0~km. Phaethon's dimensions along its principal axes are 6.4~km, 6.1~km, and 4.5~km.

The NIRSpec and NIRCam observations were taken over relatively short time periods (1.5\% and 4.1\% of the rotation period, respectively), even when including the time spent on filter changes, dithering, and readout. The observing geometries for these consecutive observations are shown in Figure \ref{fig:geom_oct}. Due to the short observing times, our observations only sample a small fraction of the surface. The NIRSpec data sample portions of the northern hemisphere and equatorial region. The observed surface latitude is $\sim$10$^{\circ}$ south of the sub-observer latitudes sampled by the December 2017 IRTF LXD spectra presented in \cite{takir2020}.

Due to the long sequence of the MIRI MRS observations, our observations covered approximately 50\% of a full Phaethon rotation period. The Spitzer observation \citep{maclennan2024} was taken over a much shorter period of time ($<$ 10~mins). Comparing the Spitzer and JWST viewing geometries (Fig. \ref{fig:geom_dec}, we see that the surface observed by Spitzer is a subset of what is seen by JWST. Figure \ref{fig:geom_dec} shows the sequence of the MIRI observation: observation start (left), observation midpoint (center), and observation end (right). The sequence shows a well illuminated surface ($\sim$5.3\% shadowed) with sub-JWST latitude of +3$^\circ$.  The amount of time required for the MIRI observation results in coverage of a large fraction of the surface of Phaethon. Therefore, our data represent an average of the spectral properties of much of the surface.

\begin{table}[ht]
\centering
\begin{tabular}{ l c c c c c }
\hline
Observation & Start Time (UT) & Effective Total & Solar & JWST & Solar Phase \\
& & Exposure Time (s)  & Distance (au) & Distance (au) & Angle ($^{\circ}$) \\
\hline
\hline
NIRSpec Fixed Slit & 2022 Oct 11 8:56:23 & 187.0 & 2.10 & 1.28 & 20.5\\
NIRCam F090W \& F430M & 2022 Oct 11 9:42:03 & 468.8  & 2.10 & 1.28 & 20.4 \\
MIRI MRS-long & 2022 Dec 10 20:24:19 & 1409.7 & 2.34 & 1.65 & 21.0 \\
MIRI MRS-medium & 2022 Dec 10 21:01:58 & 1409.7  & 2.34 & 1.65 & 21.0\\
MIRI MRS-short & 2022 Dec 10 21:39:29 & 1409.7  & 2.34 & 1.65 & 21.0\\
 \hline
\end{tabular}
\caption{JWST Observational Circumstances for Phaethon.}
\label{table:obs}
\end{table}

\section{Data Reduction} \label{sec:reduction}

\subsection{NIRCam}\label{subsec:NIRCamred}

The uncalibrated Level 1 data, consisting of stacks of non-destructive detector readouts, were downloaded from MAST and processed locally using Version 1.14.0 of the official JWST calibration pipeline \texttt{jwst} \citep{bushouse2024}; necessary reference files (e.g., flat fields and photometric calibration information determined from commissioning data and subsequent in-flight performance of the instrument) were drawn from context \texttt{jwst\_1237.pmap} of the JWST Calibration Reference Data System (CRDS). Both the pipeline version and the reference file set were the latest available resources as of 2024 April 25.

The Level 1 data were passed through all three stages of the calibration pipeline, using default pipeline parameter settings, to produce flat-fielded, bias-corrected, and flux-calibrated dither-combined images. For these and all other observations, the telescope was tracked at the non-sidereal rate of Phaethon, resulting in a well-aligned point source with no visible trailing. Prior to the second stage of the pipeline, after the ramps of non-destructive reads had been converted to countrate images, we corrected for the residual readnoise variations across the detector by subtracting the 3$\sigma$-clipped median from each row. This procedure resulted in significantly cleaner images with no discernible horizontal banding, while simultaneously removing the background flux across the field of view.

Visual inspection of the final 
dither-combined images did not reveal any obvious tail or coma in either filter. We next applied two image enhancement techniques commonly used to look for faint asymmetries around active objects \citep[e.g.,][]{Schleicher2004, Samarasinha2014}: division of a ${\rho}^{-1}$ profile (where $\rho$ is the radial distance from Phaethon) and division of an azimuthal median profile. 
Both enhancements accentuated the known artifacts due to the diffraction spikes, but did not reveal any sign of activity. We also compared the radial profiles of Phaethon in each 
dither-combined image to those of five comparison stars observed through both the F090W and F430M filters. The stars were O-, A-, and G-type absolute flux calibrator stars observed as part of Cycle 1 programs 1536, 1537, 1538 and Cycle 2 programs 4497 and 4498 (PI: Karl Gordon). The apparent $V$ magnitudes of these stars ($V{\approx}13.0$) are significantly brighter than that of Phaethon during our observations ($V=17.58$ according to JPL Horizons), so we normalized each object's radial profile by its peak brightness. Through both filters, Phaethon's radial profile was consistent with the ensemble of comparison star profiles. 

We attempted to set a quantitative upper limit to any coma that could have gone undetected. We constructed a simple model consisting of Phaethon and a coma of arbitrary brightness whose brightness decreases as ${\rho}^{-1}$, consistent with steady-state dust production \citep[cf.][]{ahearn1984}. We then convolved the model with a stellar point-spread function (PSF), measured the radial profile of the resulting image, and compared it to the radial profile we measured for Phaethon. We determined the standard deviation in each annulus used to construct the radial profile and used this as the 1-$\sigma$ uncertainty for Phaethon's brightness at that radial distance. Beyond a few pixels from Phaethon, fluctuations in the background dominated photon noise, so the annulus standard deviation yielded a more conservative upper limit to any coma present. We repeated this process for a variety of coma brightnesses, using each of the five comparison stars mentioned above, and for both the F090W and F430M filters. At the 3-$\sigma$ level, we should have been able to detect a coma whose integrated brightness within 2$\arcsec$ of Phaethon was approximately equivalent to Phaethon's brightness in the F090W filter and a few times brighter than Phaethon in the F430M filter. These are considerably less restrictive upper limits than were set by \citet{jewitt2019} using the 8-m VLT-UT3 during Phaethon's 2017 close approach to Earth due to NIRCam's bright diffraction spikes and the relatively noisy F430M comparison star images, so we do not pursue any further estimates of activity with our data. However, extrapolation to Phaethon yields a maximum undetected coma within $0\overset{''}{.}20$ of $\sim$5\% of Phaethon's signal, thus confirming that our NIRSpec and MIRI spectra are not significantly contaminated by any undetected activity.

\subsection{NIRSpec}\label{subsec:nirspecred}

The data processing and spectral extraction methodology used for the NIRSpec observations closely mirrored the techniques applied in the published analysis of fixed slit observations of the near-Earth asteroid Didymos \citep{rivkin2023}. The uncalibrated Level 1 data were passed through the first two stages of the calibration pipeline to produce flattened, dark and bias-corrected, wavelength- and flux-calibrated Level 2 spectral images for each of the two dithered exposures. The optional readnoise correction step \texttt{nsclean} was manually set to run on our data. The NIRSpec detectors exhibit correlated readnoise artifacts \citep{moseley2010,rauscher2012} that manifest as systematic flux offsets between adjacent columns. The \texttt{nsclean} routine effectively removed this vertical striping by leveraging the flux measured in the unilluminated portions of the NIRSpec 64-pixel-wide SUBS200A1 subarray that was used in our observations.

Spectral extraction was carried out using a custom empirical PSF-fitting procedure, which is described in detail in \citet{rivkin2023}. In short, the local PSF template at each column of the spectral image $j$ was estimated by median-averaging the background-subtracted cross-dispersion flux profiles in the adjacent 20 columns ($j \pm 10$). This empirical PSF template was then fit to the flux profile in column $j$ using a Levenberg-Marquardt least-squares optimization routine, with an additional additive offset parameter to simultaneously fit the background level. We then extracted the flux from the best-fit scaled PSF template using an 11-pixel-wide region centered on the centroid of the spectral trace; the centroid position was determined from a Gaussian fit to the flux profile, collapsed along the dispersion axis, and rounded to the nearest integer pixel. The background region was defined as all pixels lying more than 7 pixels from the centroid position.

\begin{figure}[ht!]
\plotone{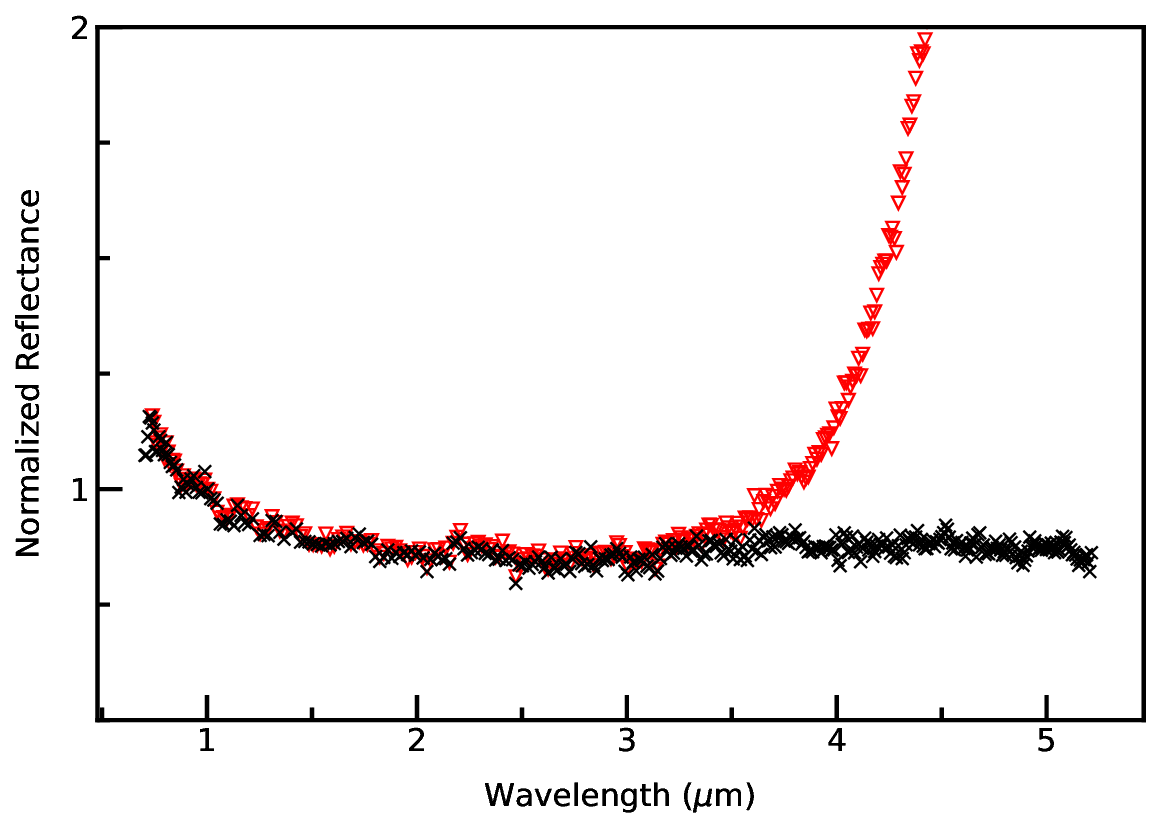}
\caption{Final NIRSpec spectrum of Phaethon with the reflected and thermal flux (red triangles) and with the thermal flux removed (black crosses) . 
\label{fig:phaethon_NIRSpec_thermal}}
\end{figure}

After extracting the spectra from both dithered exposures, each spectrum was passed through a 21-pixel-wide $3\sigma$ moving median filter to remove outliers. Then, the two spectra were averaged together to produce the final combined irradiance spectrum of Phaethon. In order to calculate the reflectance spectrum, account for flux losses outside of the extraction aperture, and remove any residual instrumental systematics, we applied the same data processing and spectral extraction procedure on NIRSpec fixed slit observations of the solar standard star P330-E, obtained as part of flux calibration program 1538 (PI: Karl Gordon). We utilized the same spectral extraction box size and divided the resultant combined stellar spectrum from Phaethon's irradiance spectrum to arrive at the final reflectance spectrum, which is plotted in Figure \ref{fig:phaethon_NIRSpec_thermal}. 

The thermal tail correction involves determining the relative amounts of reflected and emitted light as a function of wavelength, and then removing the emitted component. Wavelengths with negligible thermal emission remain unchanged. 
The thermal correction was done using a modified version of the Standard Thermal Model \citep[STM;][]{lebofsky1986refined} that treats the beaming parameter ($\eta$) as a free parameter \citep[as done in the Near Earth Asteroid Thermal Model (NEATM),][]{harris1998neatm}. We do not anticipate a large contribution from the night side given that our observations were taken at a small phase angle ($\alpha$, Fig. \ref{table:obs}). The beaming parameter is the only free parameter in this implementation of the modified STM because all other parameters are known from previous publications \citep[e.g.,][]{hanus2016,kareta2018,taylor2019}, from the observing circumstances (i.e., $r_h$, distance from JWST $\Delta$, phase angle $\alpha$), or can be assumed and fixed (emissivity=0.9, slope parameter G=0.15). Because the ratio of thermal and reflected light is used, the thermal tail removal is insensitive to the specific choice of diameter and G values. Different diameter and G values merely result in different best-fit beaming parameters. While the beaming parameter is associated with thermal properties \citep[e.g.,][]{harris2016thermal}, we do not use it in this work. As an example, if we set G=0, the change in reflectance throughout the spectrum is $<$ 1\%. 
At Phaethon's solar distance during these observations, thermal emission was less than 2\% of the total emission at wavelengths shortward of 3.35 $\mu$m, and so the spectral behavior and spectral slope are basically unaffected by the thermal model. We adjusted the beaming parameter such that the resulting thermal emission model removes the thermal flux longward of 3.35 $\mu$m to give the entire dataset an overall slope of zero.
This zero slope is consistent with the \cite{takir2020} analysis in this wavelength range. The original spectrum and the spectrum with the thermal flux removed can be seen in Figure \ref{fig:phaethon_NIRSpec_thermal}.

Some structure can be seen in the resulting spectrum. For wavelengths between 3.15 and 3.85~$\mu$m, the reflectance is up to 2\% higher than the mean value, with a peak near 3.7~$\mu$m. Additionally, the reflectance at 2.4--3.15~$\mu$m is lower than the zero-slope mean value (with no robustly identifiable minimum). We do not interpret this lower reflectance region as an absorption band resulting from hydrated minerals (see Section 4.1). 

\subsection{MIRI}\label{subsec:mirired}

Data processing and spectral extraction of the MIRI MRS observations proceeded in much the same way as for the NIRSpec fixed slit observations. The uncalibrated Level 1 data were passed through the first two stages of the JWST calibration pipeline, resulting in fully-calibrated IFU data cubes for each of the four dithered exposures. Each slice within the data cubes corresponds to a resampled image of the field of view at a particular wavelength, with pixel scales of $0\overset{''}{.}13$ and $0\overset{''}{.}20$ in spectral channels 1--2 and 3--4, respectively. Data collected by the MIRI MRS are characterized by strong spectral fringing, particularly at the short- and long-wavelength ends of each sub-band \citep{argyriou2020}. The calibration pipeline contains several routines that work to reduce the effect of the fringes on the extracted spectra: (1) the \texttt{fringe} step in Stage 2, which utilizes a dedicated fringe flat, (2) the optional \texttt{residual\_fringe} step in Stage 2, which applies a periodogram analysis to the detector image and was manually turned on when processing our data, and (3) a final spectrum-level defringing routine included as part of Stage 3 of the calibration pipeline, which we applied to each sub-band spectrum following spectral extraction.

A two-dimensional version of the empirical PSF fitting method described in the previous subsection was applied to the MIRI MRS data cubes. This implementation is based on the technique that was developed in previous works analyzing NIRSpec IFU data of faint solar system objects \citep[e.g.,][]{emery2024,grundy2024,wong2024}. The calibration pipeline outputs data cubes in surface brightness units (MJy/sr), and we converted the pixel values to flux units (Jy) by multiplying the data cubes by the pixel area listed in the file headers prior to spectral extraction. The PSF templates were generated using an 11-slice-wide moving window along the wavelength axis; the narrower range was chosen to prevent biases in the PSF template generation induced by the remaining fringing artifacts present in the wavelength slices. The extraction region was defined as a 7$\times$7 pixel box centered on the calculated centroid position of Phaethon; the background region included all pixels located outside of a 31$\times$31 box surrounding the centroid. After running the individual extracted sub-band irradiance spectra through the final defringing procedure, 3$\sigma$ outliers were masked following the same method used on the NIRSpec spectra, before the groups of four dithered spectra were averaged together. The amount of fringing remaining in the spectra is at the 1\% level.

The size of a point-source PSF in the MIRI MRS wavelength cubes is quite large, particularly at the longest wavelengths, where it can fill up the majority of the field of view. To correct for wavelength-dependent flux losses outside of our fixed aperture due to the growing source PSF, we followed the technique outlined in \citet{rivkin2023} and empirically derived flux correction factors using Cycle 1 observations of the A-type standard star del Umi, obtained as part of flux calibration program 1536 (PI: Karl Gordon). After extracting the star's irradiance spectrum using the same aperture size and background region as in the case of Phaethon, each sub-band's spectrum was divided by the CALSPEC model spectrum of del Umi \citep{bohlin2014}. Cubic polynomial functions were then fit through the resultant ratio arrays to generate smooth correction vectors for all sub-bands, which were then applied to our extracted spectrum of Phaethon to recover the full source flux at every wavelength.

At this stage, significant offsets remained between adjacent sub-bands across the spectrum. As described in Section~\ref{sec:obs}, the MIRI observation sequence spanned a significant fraction of Phaethon's rotation period, and we attribute these systematic offsets to rotational variations in Phaethon's overall brightness between observations obtained with the different grating settings. We kept the sixth sub-band spectrum (corresponding to spectral channel 2 and the long grating setting) in place and rescaled the other sub-bands by matching the average fluxes in adjacent sub-bands within the overlapping wavelength regions. This sub-band was used for scaling because it has high signal-to-noise and it is in the middle of the spectrum. 

We converted the thermal emission spectrum to an emissivity spectrum using the NEATM code \texttt{mskpy} \citep{Kelley2021mskpy}. The model calculates the observed flux by integrating the predicted thermal emission over the visible hemisphere of the target. For the Phaethon MIRI observation sequence, we used the observing geometry ($r_h$, $\Delta$, and $\alpha$) with diameter $D$=5.1~km and visible geometric albedo $p_v$=0.122 \citep{hanus2016}. We did not solve for diameter and albedo independently and used the published values to set the initial fit parameters. The \texttt{mskpy} implementation of NEATM also includes the beaming parameter ($\eta$) and emissivity ($\epsilon$). We used the mean emissivity for asteroids of $\epsilon$=0.9 from \cite{mainzer2011}. The \texttt{mskpy} implementation uses a minimization $\chi^2$, which allows the diameter and beaming parameter to vary in the fit process.  We do not include the NEATM-derived diameter or beaming parameter in our analysis. We convert flux to emissivity by dividing the measured flux by the modeled flux. We use standard error propagation for our formal error on the emissivity. We limit our analysis to MIRI data with $\lambda>7~\mu$m due to low SNR and intrinsic flux for shorter wavelengths.

\begin{figure}[ht!]
\plotone{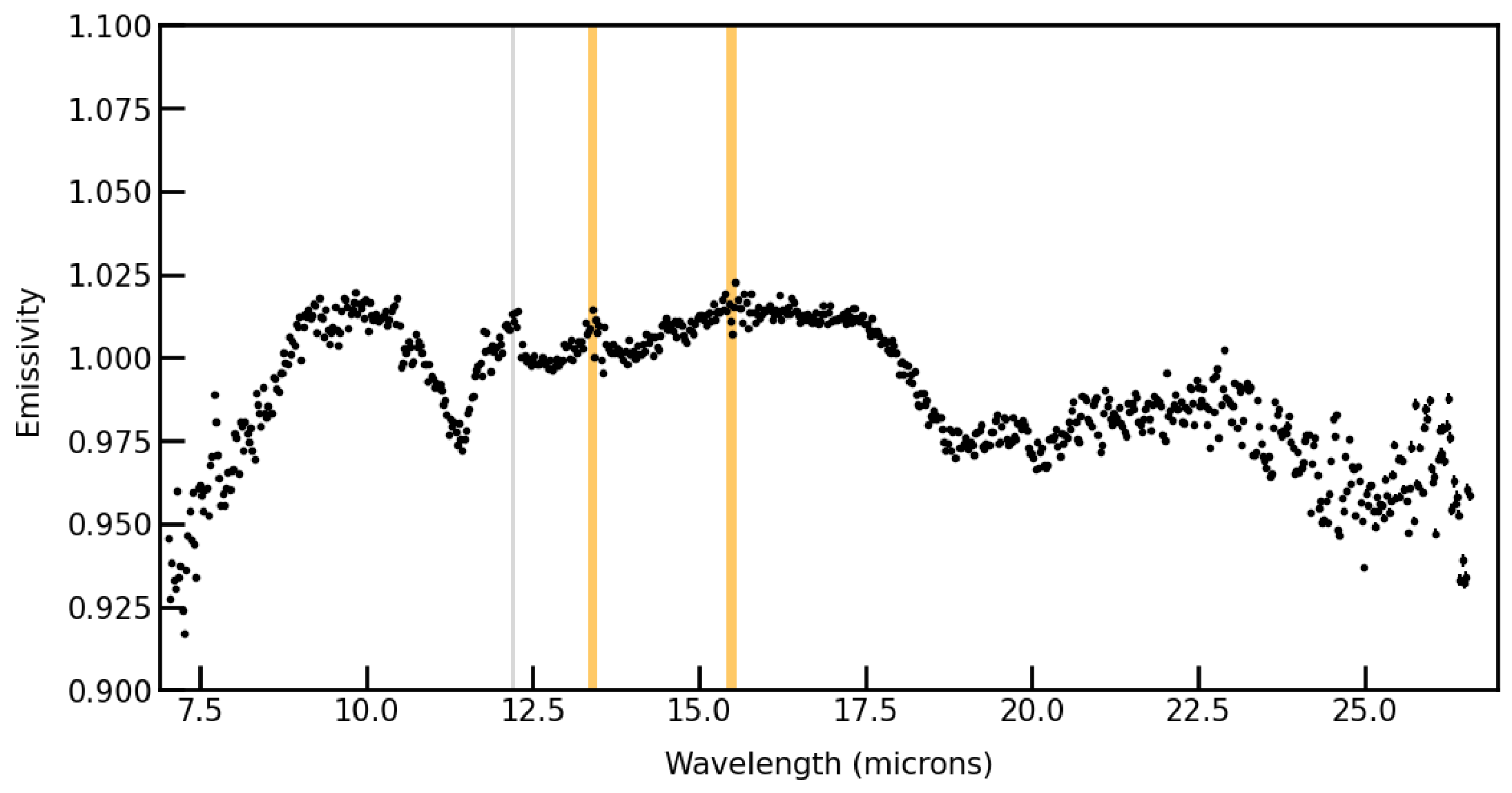}
\caption{Final MIRI emissivity spectrum of Phaethon. We note the documented spectral leak at $\sim$12.2 $\mu$m as a gray line and the edges of sub-band 3B as gold bands. We use this artifact labeling methodology in subsequent figures.
\label{fig:miri}}
\end{figure}

An initial investigation of the nominal emissivity spectrum showed local maxima at $\sim$13.5 and 15.5 $\mu$m, which correlate with the edges of sub-band 3B (channel 3, medium), and a slight overall offset of the sub-band compared to the adjacent sub-bands. The pipeline defringing routines are not perfect, and significant residual fringing is apparent at the short- and long-wavelength edges of some sub-bands. The location of these maxima at the edges of sub-band 3B, along with the discernible increase in scatter, means that the maxima are likely not due to the composition or the physical properties of the target, but rather stem from the biasing effect of residual fringing on the previously described sub-band renormalization process. In the final spectrum (Fig.\ \ref{fig:miri}), we rescaled the sub-band 3B segment by a factor of 0.995 relative to the nominal scaling to reduce the impact of the flux offset on the final spectrum. All spectral figures demarcate the edges of sub-band 3B as gold bands and the documented spectral leak around 12.2 $\mu$m\footnote{See \url{https://jwst-docs.stsci.edu/known-issues-with-jwst-data/miri-known-issues/miri-mrs-known-issues}} as a gray line. We caution against the use of these apparent ``features" in any spectral analysis.

\section{Results} \label{sec:results}

\subsection{NIRSpec}\label{subsec:nirspecres}

\begin{figure}[ht!]
\plotone{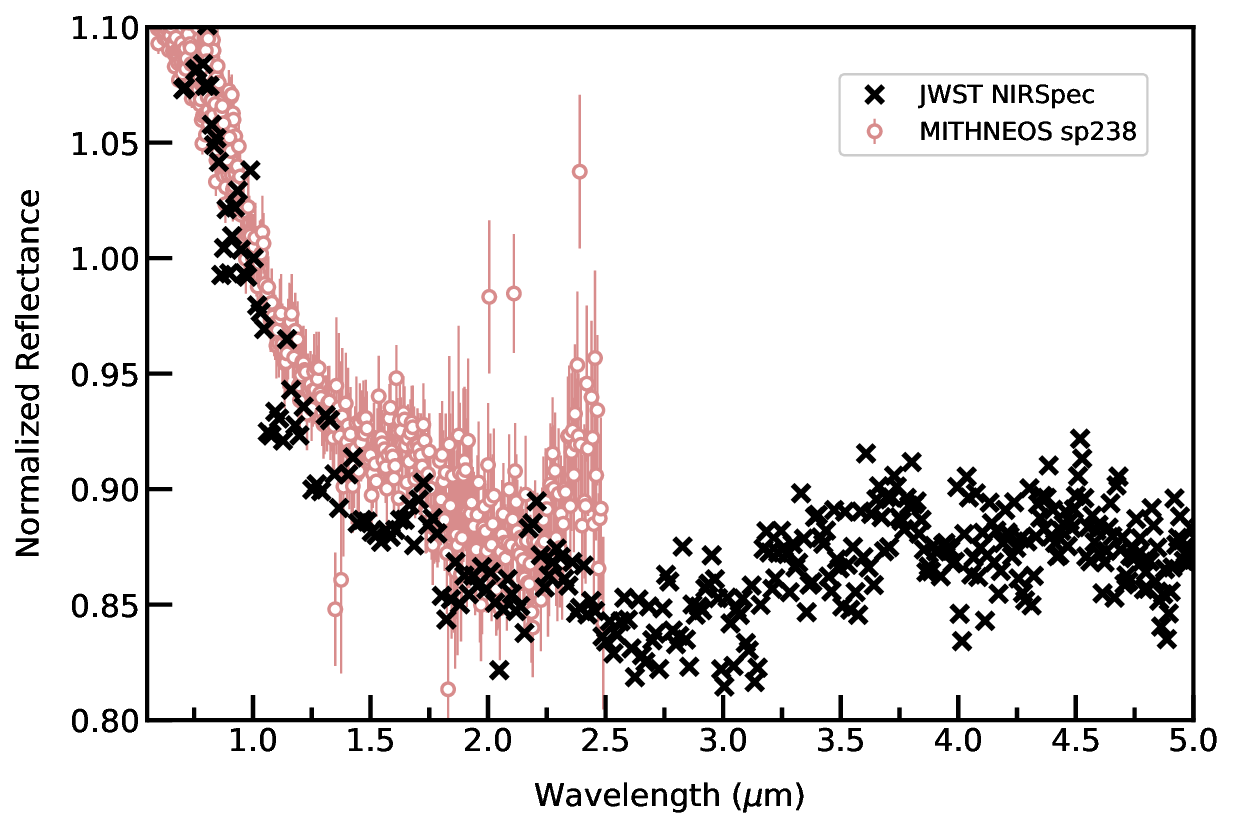}
\caption{Final NIRSpec reflectance spectrum of Phaethon. We compare the data to a ground-based near-infrared spectrum from the MITHNEOS survey \citep[run ``sp238",][]{binzel2019compositional}. Both spectra are normalized to unity at 1.0 $\mu$m. The two spectra show similar slopes and spectral shape in their common wavelength regions. }
\label{fig:phaethon_mithneos}
\end{figure}

\begin{figure}[ht!]
\plotone{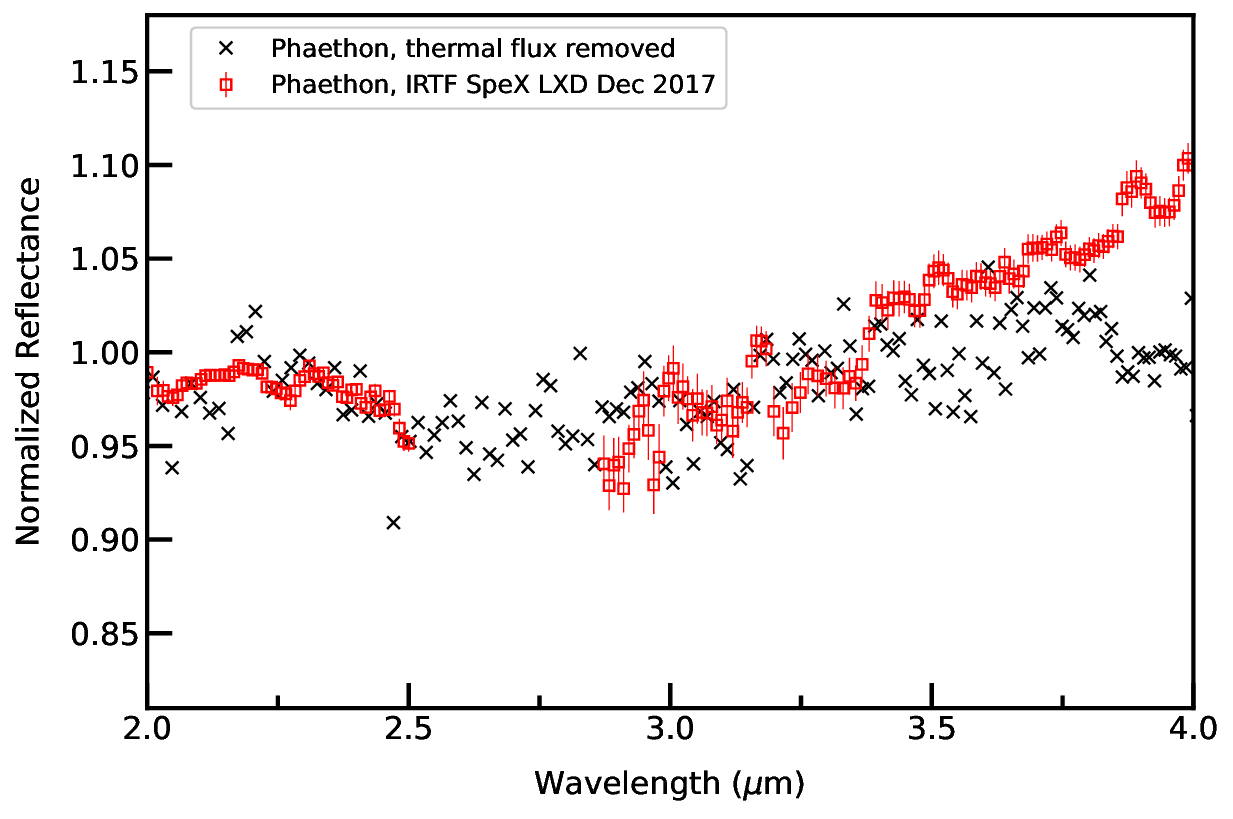}
\caption{We compare the 3-$\mu$m region of our NIRSpec data to a ground-based observation from \cite{takir2020}. We confirm the lack of absorption band in these wavelengths. 
\label{fig:phaethon-takir}}
\end{figure}

The final NIRSpec reflectance spectrum of Phaethon is shown in Figure \ref{fig:phaethon_mithneos} compared to a ground-based near-infrared spectrum from the MITHNEOS survey \citep[e.g.][]{binzel2019compositional}. The spectra are consistent in their common wavelength regions. The MITHNEOS survey uses the SpeX instrument on the NASA Infrared Telescope Facility \citep{rayner2003spex}. The figure contains the spectrum from run ``sp238" (11 November 2017), which is consistent with the other MITHNEOS spectra of Phaethon (see Section 5). We also compare the JWST NIRSpec data to previous observations of Phaethon in the 3-$\mu$m region taken from the IRTF \citep[SpeX LXD mode from][Figure \ref{fig:phaethon-takir}]{takir2020}. The spectra largely agree, and we attribute the differences at the longest wavelengths to different choices of thermal model between our work and that of \cite{takir2020}.  

\begin{figure}[ht!]
\plotone{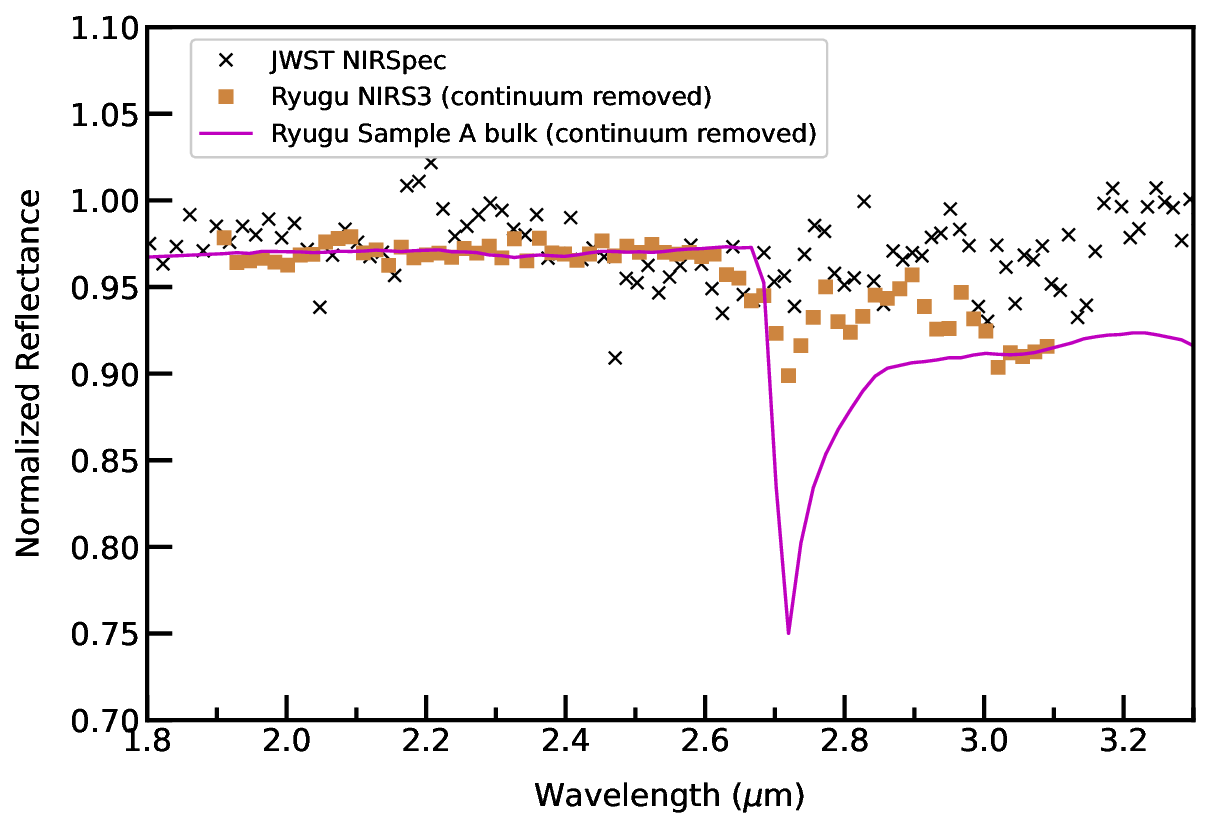}
\caption{We compare the 3-$\mu$m region of our NIRSpec data to Ryugu using in situ Hayabusa2 observations \citep{matsuoka2023space} and laboratory data of the returned sample \citep{nakamura2022formation}. Ryugu is clearly more hydrated than Phaethon. 
\label{fig:phaethon_ryugu}}
\end{figure}

The NIRSpec data show no hint of an absorption band in the 2.5-2.85-$\mu$m region. Quantification of the band depth is limited by the scatter in the data. We estimate a band depth of less than 2\% at 2.7-$\mu$m. \cite{beck2021water} determined a relationship between the hydrogen content of carbonaceous chondrite phyllosilicates and their band depths at 2.75 and 2.80-$\mu$m. Adopting the 2\% upper limit for the band depths at those wavelengths results in a conservative estimate of an upper limit on hydrogen content by mass of 0.06\% in phyllosilicates within Phaethon's regolith. For comparison, \cite{vacher2020hydrogen} found a range of hydrogen content of 0.07-0.16\% in CV chondrites, which included hydrogen in all CV minerals, including organic materials. Our estimated upper limit suggests an anhydrous surface for Phaethon. Figure~\ref{fig:phaethon_ryugu} compares Phaethon's spectrum to those of Ryugu \citep[in situ and the returned surface sample;][]{matsuoka2023space}. The absorption bands in both Ryugu spectra are clearly deeper than any discernible features in the Phaethon spectrum.

\subsection{MIRI}\label{subsec:mirires}

\begin{figure}[ht!]
\plotone{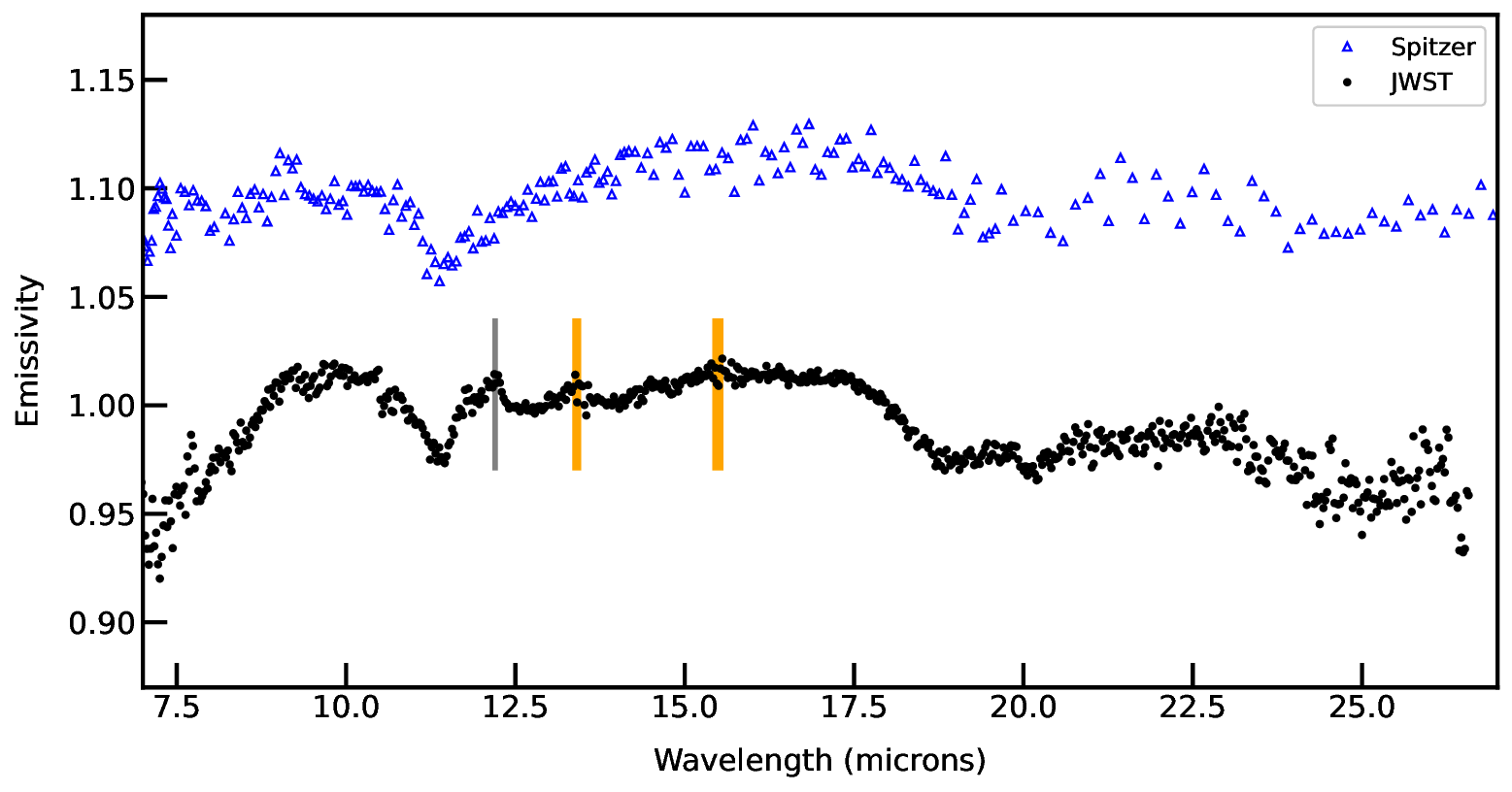}
\caption{The emissivity spectra from JWST MIRI and Spitzer IRS are largely consistent with each other. The JWST data shows higher signal-to-noise, which enables a clearer view of diagnostic spectral features at $\sim$19-20.2~$\mu$m and $\sim$15-17~$\mu$m. 
\label{fig:spitzer}}
\end{figure}

Our final Phaethon emissivity spectrum is consistent with the previously published Spitzer IRS spectrum \citep[i.e.,][Fig. \ref{fig:spitzer}]{maclennan2024}. The MIRI spectrum shows higher signal-to-noise,  especially for $\lambda>18~\mu$m, which enables a more detailed investigation of the spectral features. We use two spectral regions to further constrain the composition and meteorite analog: $\sim$19-20.2~$\mu$m and $\sim$15-17~$\mu$m.

\begin{figure}[ht!]
\plotone{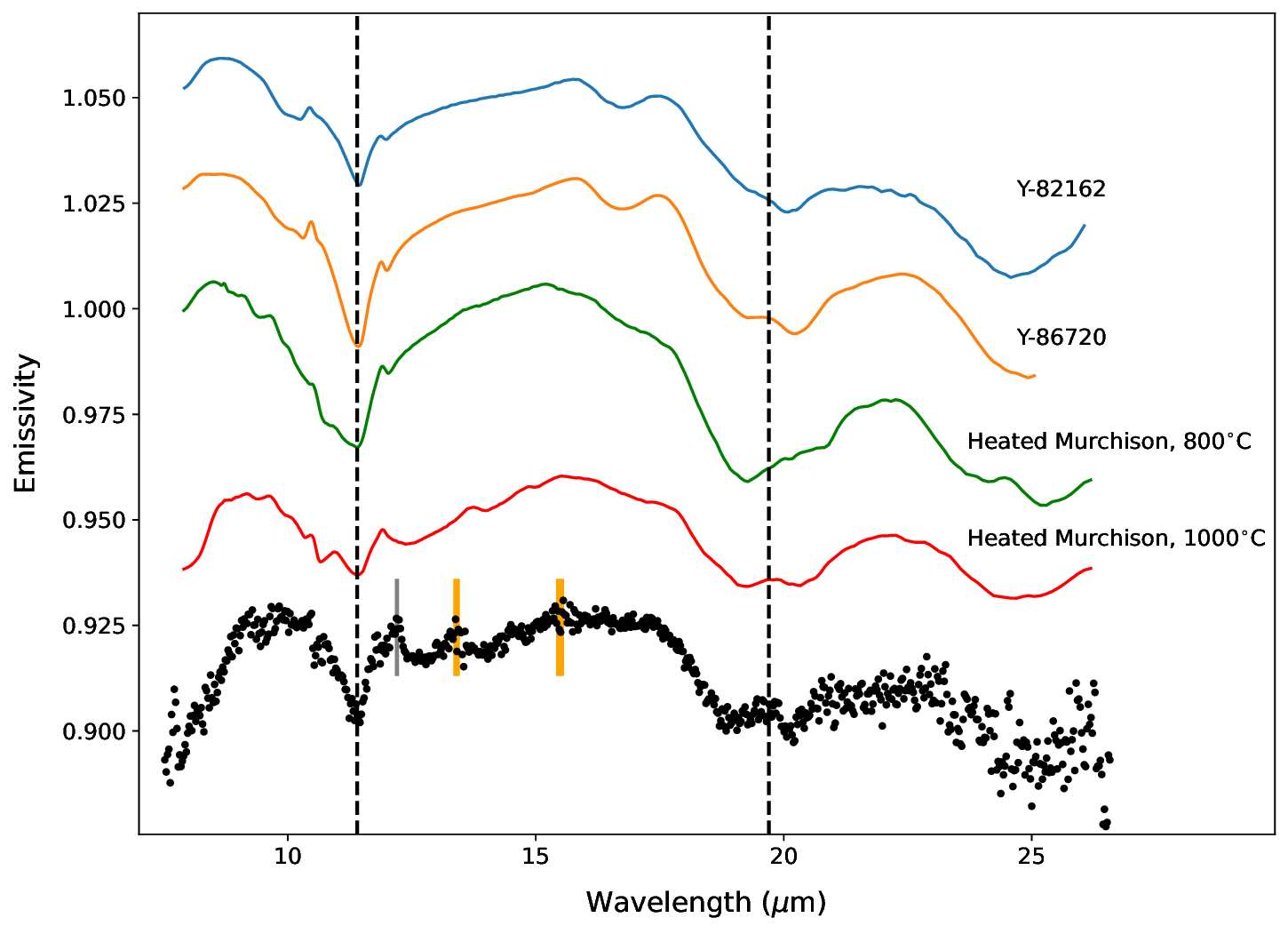}
\caption{Our analysis shows that Phaethon is most similar to the CM Murchison sample heated to 1000$^{\circ}$~C. For clarity on the positions of the spectral features, the spectra for Phaethon and Y-86720 are scaled by a factor of 0.6 and all other spectra are scaled by a factor of 1.3 before being offset in emissivity.
\label{fig:CMCY}}
\end{figure}

Our spectrum shows additional spectral structure at $\sim$19-20.2~$\mu$m that is not distinguishable in the Spitzer data. The local maximum seen in the MIRI data at 19.7~$\mu$m is not present in the Y-82162 spectrum that \cite{maclennan2024} determined to be the best spectral analog (Fig. \ref{fig:CMCY}). \cite{king2019yamato} describe the Y-82162 meteorite as being an aqueously altered object that subsequently experienced short-lived thermal metamorphism with temperatures of at least 500$^{\circ}$~C ($\sim$773~K). This peak temperature is notably lower than the maximum temperature experienced by Phaethon at perihelion (i.e., T= 1050~K). \cite{maclennan2024} also noted that a laboratory heated sample of Murchison was a good spectral match to Phaethon, but it was eliminated as a spectral analog due to a mismatch of spectral features at 21~$\mu$m (see their Fig. 2). We used RELAB \citep{milliken2020relab} mid-infrared spectra of heated Murchison samples, Y-82162, and Y-86720 to re-examine potential analogs (Figure \ref{fig:CMCY}, Table \ref{table:relab} for the specimen IDs). These specific CY meteorites have been suggested as analogs for Phaethon by \cite{lazzarin2019phaethon} and \cite{licandro2007nature}, respectively. Additionaly, \cite{king2019yamato} concluded that the Y-86720 meteorite had reached temperatures in excess of 700$^{\circ}$~C, which is more comparable to the peak Phaethon temperature than Y-82162. We attempt to mitigate the impacts of grain size on the spectral comparison by selecting the best grain size available for each sample. 
\cite{devogele2020new} calculated a regolith grain size of 3-30 mm for Phaethon using the method by \cite{gundlach2013new} and a thermal inertia derived from NEOWISE data \citep{mainzer2014initial}. The derived grain sizes are larger than those available in the RELAB collection, so we use the largest grain size available (75-125~$\mu$m) for the Murchison data. We also select the largest grain size available (63-125~$\mu$m) for our spectral comparison of Y-82162. The only mid-infrared spectrum available for Y-86720 was of a meteorite chip. 
A comparison of two heated Murchison spectra (at 800$^{\circ}$~C \& 1000$^{\circ}$~C for a week) and the CY spectra to the Phaethon emissivity spectrum shows that the structure of the $\sim$19-20.2~$\mu$m wavelength region is most similar to that of the 1000$^{\circ}$~C heated Murchison and Y-86720. This spectral similarity demonstrates that this feature on Phaethon is likely the result of thermal alteration of previously aqueously altered material. 

The MIRI spectrum shows no band in the $\sim$15-17~$\mu$m region. Laboratory investigations \citep[e.g.,][]{hanna2020distinguishing} show an increase in the depth of the Mg-OH band ($\sim$16~$\mu$m) with an increasing proportion of Mg-Fe phyllosilicates relative to anhydrous silicates. The lack of a feature here suggests a lack of hydration in the surface regolith. We note that a feature in this region could also be due to magnesium-rich olivine produced during the dehydration process. Therefore, the presence of this feature for Y-86720 could be indicative of anhydrous olivine and not phyllosilicates. Either way, the CY meteorites are not the best spectral match to Phaethon given the improved SNR of the MIRI data.

We conclude that Phaethon is most similar to the 1000$^{\circ}$~C heated CM Murchison meteorite. The spectral similarity extends beyond the $\sim$15-17~$\mu$m and $\sim$19-20.2~$\mu$m to include the 10~$\mu$m emissivity plateau and the spectral shape for wavelengths greater than 20~$\mu$m. 

Given the similarity of the MIRI and Spitzer spectra and the range of viewing geometries within the MIRI observation (Fig. \ref{fig:geom_dec}), the data suggest that the spectrum is consistent across the surface of the object. This homogeneity is supported by the lack of notable spectral variability seen in \cite{kareta2018} and \cite{takir2020}. Due to our single observation at an equatorial observing geometry, we cannot use the data to examine the \cite{maclennan2022surfacehetero} conclusion that the northern and southern hemispheres have different thermophysical properties potentially due to regolith grain size and surface structure or investigate the change in thermal inertia with respect to northern hemisphere latitude observed by \cite{2021DPSVervack}. 

The heated Murchison spectra show a local maximum around 12.2~$\mu$m. This is close to the position of the known light leak artifact. It is possible that the strength of the emissivity bump at 12.2~$\mu$m is due to the light leak occuring over a true spectral feature. 

\begin{table}[ht]
\centering
\begin{tabular}{ l l }
\hline
Specimen ID & Name \& Description \\

\hline
\hline
MB-TXH-064-D4 & CM2 Murchison, 75-125 $\mu$m\\
MB-TXH-064-E4 & CM2 Murchison heated at 400~C, 75-125 $\mu$m \\
MB-TXH-064-G4 & CM2 Murchison heated at 600~C, 75-125 $\mu$m \\
MB-TXH-064-I4 & CM2 Murchison heated at 800~C, 75-125 $\mu$m \\
MB-TXH-064-K4 & CM2 Murchison heated at 1000~C, 75-125 $\mu$m \\
MB-CMP-019-C & Y-82162, 63-125 $\mu$m\\
MP-TXH-159 & Y-86720, chip\\
\hline
\end{tabular}
\caption{RELAB mid-infrared spectra for heated samples of Murchison.}
\label{table:relab}
\end{table}

\begin{figure}[ht!]
\plotone{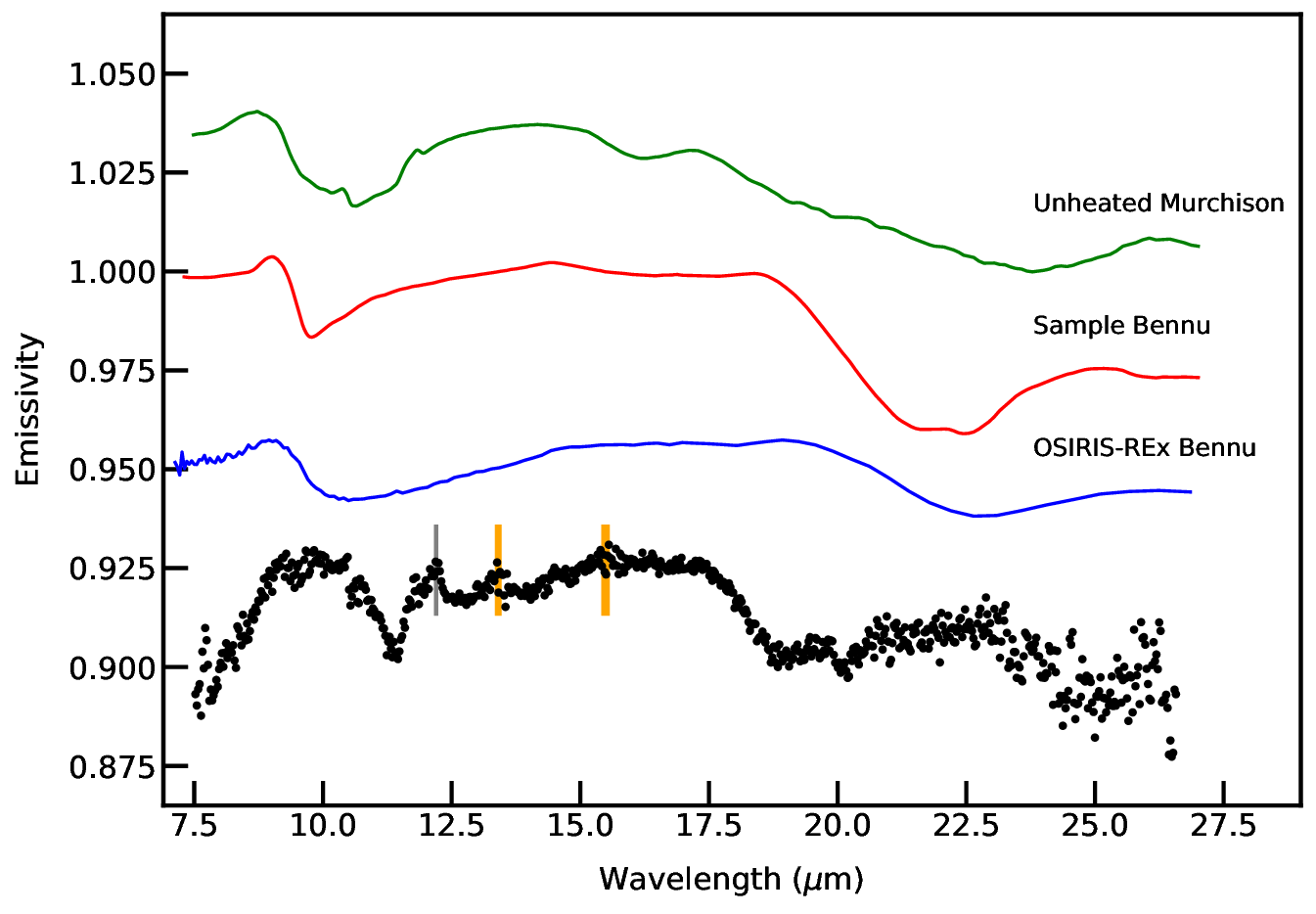}
\caption{The mid-infrared spectra of B-types Phaethon and Bennu are distinctly different from each other with key features at different wavelengths and distinct spectral structure. Both spectra are also different from unheated (and heated) samples of Murchison. For clarity, Phaethon is scaled by a factor of 0.6 and all spectra are offset from Phaethon in emissivity.
\label{fig:bennu}}
\end{figure}

\section{Discussion} \label{sec:disc}

From the current orbit of Phaethon, \cite{maclennan2021dynamical} determined the maximum surface temperature of the object at perihelion to be T=1050~K $(\sim$777$^{\circ}$~C). This is a couple of hundred degrees cooler than the 1000$^{\circ}$~C (1273~K) temperature of the heated Murchison sample that presents the best spectral match. Figure \ref{fig:CMCY} shows that the CM spectrum with the best temperature match (800$^{\circ}$~C) does not show the distinct feature at 19.7~$\mu$m that we see in the Phaethon data. If the surface remains dehydrated \citep{maclennan2024}, then the current JWST spectra will show the outcome of the dehydration process. Our heated CM meteorite analog suggests that it is possible that the surface of Phaethon was heated beyond 1050~K in the past or that the amount of time that Phaethon has spent at the highest temperatures ($>$ 1 week, as done in the laboratory) has further altered the surface beyond what is seen in the 800$^{\circ}$~C RELAB spectrum.  

Phaethon made 13 perihelion passes in the time between the January 2005 Spitzer observation and the JWST observations presented in this work. 
The spectral consistency over time suggests that the recent close solar approaches are not actively changing the spectral characteristics of the surface. This consistency supports the \cite{maclennan2024} hypothesis that the surface is dehydrated and remains dehydrated with any rehydration occurring in the sub-surface layers. The surface of Phaethon is unlikely to dramatically change unless it is heated further or subsurface layers are revealed. It is possible that the surface, and resulting spectrum, are still undergoing a slower evolution that cannot be detected over the period of time between the Spitzer and JWST observations. The lack of spectral variability does not necessarily contradict \cite{maclennan2022surfacehetero}'s conclusion that Phaethon's northern and southern hemispheres have different grain sizes. It is possible that more direct observations of the poles of Phaethon will result in different spectral properties because of the effects of grain size on the spectra. These observations have not been obtained from ground and space-based telescopes since we are limited by the sub-observer latitudes.

\cite{maclennan2021dynamical} performed an analysis of surface temperatures over an entire Phaethon orbit as a function of latitude. They calculated maximum surface temperatures for the current spin pole of Phaethon (estimated as a solstice point $\lambda_{\nu}\sim$270$^{\circ}$). The peak temperature (1050~K, noted previously) is found for equatorial latitudes, which are where both Spitzer and JWST MIRI observed. \cite{maclennan2021dynamical} found that the maximum temperatures at the poles was $\sim$760~K. The entire surface has therefore been subject to high temperatures, but the level of thermal alteration may vary over the surface. 

We compare the mid-infrared spectrum of Phaethon with the other well-studied B-type \citep{clark2011asteroid} near-Earth asteroid, Bennu, in Fig. \ref{fig:bennu}. The spectra of the two are notably different from each other with key features at different wavelengths and with a different overall structure. The Bennu spectrum is also distinct from the collection of heated Murchison spectra (including those at lower temperatures), which show key spectral features at similar wavelengths throughout the dataset. We note that \cite{rozitis2020implications} determined Bennu's maximum surface temperature at perihelion (for an equatorial surface element and zero tilt) to be 390~K ($\sim$117~$^{\circ}$C), which is lower than any of the temperatures in the heated Murchison dataset. The lower surface temperatures experienced by Bennu is supported by the analysis of \cite{delbo2011temperature}, who used dynamical evolution and thermal models to determine that there was a $<$ 20\% probability that Bennu's surface reached temperatures $>$ 600$^{\circ}$C. A comparison of Bennu to the unheated Murchison spectrum shows a more qualitatively similar spectral shape than is seen with the heated data. However, the comparison of Bennu to the unheated Murchison is far from a perfect match. There are notable differences in some spectral shapes and the wavelengths at which they occur. This is consistent with the \cite{lauretta2024asteroid} analysis of the Bennu sample, which found that the sample elemental abundances are a better match to the CI chondrites than the CMs. Therefore, it follows that Bennu and Phaethon had similar, but not necessarily the same, starting surface compositions.

\begin{figure}[ht!]
\plotone{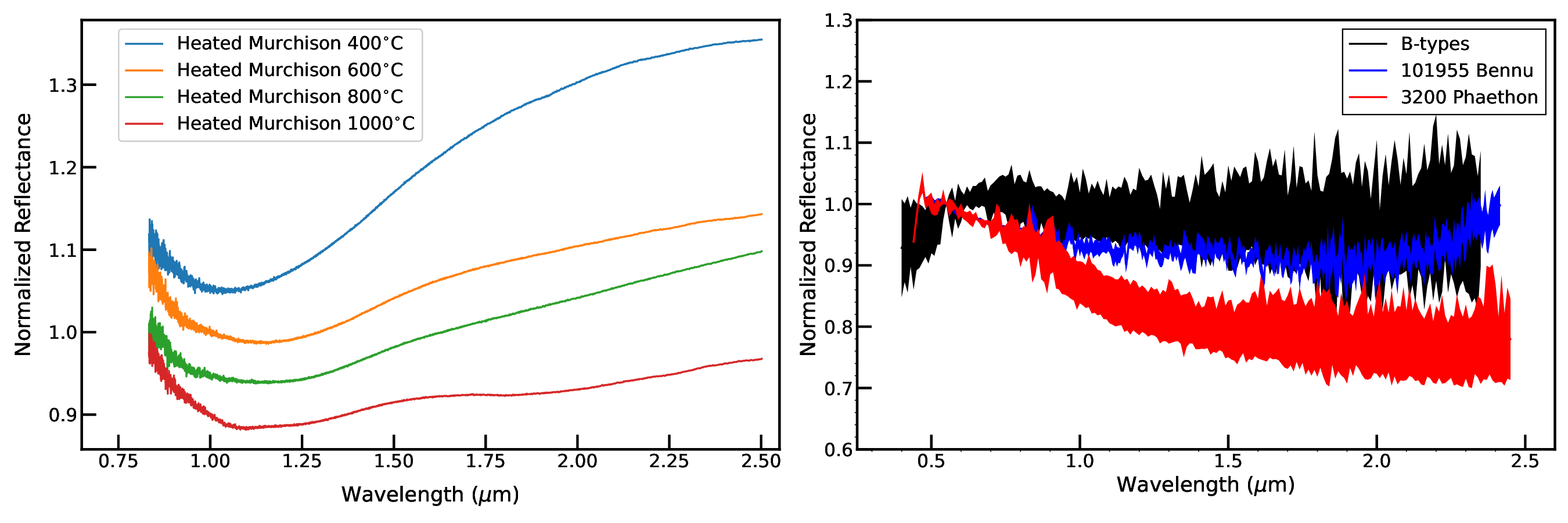}
\caption{Left: Visible and near-infrared spectra from RELAB show that the observed spectral slope of CM chondrite Murchison decreases significantly with increased temperature. Right: The visible and near-infrared spectrum of Phaethon has a much bluer spectral slope than Bennu and other B-type near-Earth objects.  
\label{fig:VNIR}}
\end{figure}

The blue spectral slope of Phaethon is likely due to a combination of the intense heating to which the object has been exposed during perihelion passages and the large regolith grain size. Figure \ref{fig:VNIR} compares the visible and near-infrared spectrum of Phaethon from the MITHNEOS program \citep[e.g.,][]{binzel2019compositional} with other B-types and a sequence of heated Murchison samples. Phaethon shows a notably bluer slope compared to Bennu and the other classified B-types. It is clear from the RELAB data that the observed spectral slope of Murchison decreases significantly with increasing temperature. Laboratory investigations of Murchison in \cite{binzel2015spectral} also show that increasing grain size leads to bluer sloped spectra. As noted previously, \cite{devogele2020new} calculate a regolith grain size of 3-30 mm for the object, which is larger than the grain sizes in the Binzel work. The contribution of intense heating to the blue slope of Phaethon is supported by observations of other low-perihelion objects.
Blue colors have been measured at visible wavelengths for other objects having perihelion closer to the Sun than Phaethon including the nuclei of comet 96P/Machholz~1 \citep{eisner2019}, ambiguous comet/asteroid objects 322P/SOHO \citep{knight2016} and 323P/SOHO \citep{hui2022}, and numerous asteroids \citep{holt2022}. Their study with JWST could enable further understanding of the extent to which extreme heating affects asteroidal surfaces.

The connection between intense solar heating and Phaethon's blue spectral slope further supports the argument that Phaethon is not from Pallas \citep[e.g.][]{kareta2018,maclennan2021dynamical}. A dynamical analysis by \cite{maclennan2021dynamical}, using the methodology of the \cite{granvik2018debiased} model, suggests that Phaethon was unlikely to have been on a Pallas-like orbit while in the Main Belt. They argue that an inner asteroid belt source with $i<$10$^{\circ}$ is most likely. \cite{granvik2018debiased} gives the probability of Phaethon being delivered to near-Earth space through the $\nu_6$ secular resonance as $\sim$64\%. \cite{maclennan2021dynamical} notes that the C-type Svea family could be a source given its location and that a notable fraction of the family are classified as B-type \citep{morate2019last}. Our spectral slope analysis suggests that the most appropriate main-belt analog family should be composed of hydrated C-complex (i.e., Ch, Cgh) asteroids. 

The DESTINY\textsuperscript{+} mission \citep{ozaki2022mission} will encounter Phaethon in 2028. Their visible and near-infrared wavelength imager will enable detailed studies of the surface, including any variation in the spectrophotometric slope. Notable variations in the surface color could indicate potential surface alteration. \cite{hanus2016} modeled Phaethon's obliquity over time (from their preferred pole solution: 319$^{\circ}$, -39$^{\circ}$) and determined that both the northern and southern hemispheres have been irradiated at perihelion in the past 50 kyr of evolution. The most current pole solution (used for Figures \ref{fig:geom_oct} and \ref{fig:geom_dec}) is 14$^{\circ}$ away at 318$^{\circ}$, -53$^{\circ}$.  Given the dynamical evolution, it is likely that all parts of the surface would have been exposed to high levels of solar heating. For its current orbit, \cite{maclennan2021dynamical} conclude that the maximum global temperature occurs at equatorial latitudes and that the maximum temperature at 1 thermal skin depth ($l=2.56$~cm) falls to $\sim$400 K. If any sub-surface material has been exposed due to impacts or as a result of the emission, it would likely have different spectrophotometric properties than the majority of the surface.

\section{Conclusions} \label{sec:conc}

We analyzed JWST NIRCam, NIRSpec, and MIRI observations to further constrain the composition of Phaethon's surface. We conclude that the surface is currently dehydrated following thermal alteration caused by the high temperatures experienced near perihelion. Our NIRSpec data lacks a 3-$\mu$m absorption feature, which supports the results of previous work by \cite{takir2020}. From the MIRI spectrum, we conclude that Phaethon's surface composition aligns more closely with heated CM Murchison samples than with the CY meteorites proposed as analogs by \cite{maclennan2024}. The MIRI spectrum matches the previous Spitzer observation, which indicates no apparent surface changes despite multiple perihelion passages. This study advances our understanding of Phaethon's surface and provides additional information for the upcoming DESTINY\textsuperscript{+} mission.

\section*{Acknowledgements}
\vspace{1em}
\noindent This work is based on observations made with the NASA/ESA/CSA James Webb Space Telescope. The data were obtained from the Mikulski Archive for Space Telescopes at the Space Telescope Science Institute, which is operated by the Association of Universities for Research in Astronomy, Inc., under NASA contract NAS 5-03127 for JWST. These observations are associated with program \#1245.


Support for program AR-2537 was provided by NASA through a grant from the Space Telescope Science Institute, which is operated by the Association of Universities for Research in Astronomy, Inc., under NASA contract NAS 5-03127. HBH and SNM also acknowledge support from NASA JWST Interdisciplinary Scientist grant 21-SMDSS21-0013.

This research utilizes spectra acquired by Takahiro Hiroi and Carle Pieters with the NASA RELAB facility at Brown University.

%

\vspace{5mm}
\facilities{JWST}
\software{\texttt{astropy} \citep{astropy2013, astropy2018, astropy2022}, \texttt{jwst} \citep{bushouse2024}, \texttt{matplotlib} \citep{hunter2007}, \texttt{numpy} \citep{harris2020}, \texttt{mskpy} \citep{Kelley2021mskpy} and \texttt{scipy} \citep{virtanen2020}.}





\bibliography{phaethon}{}
\bibliographystyle{aasjournal}



\end{document}